\documentclass[pra,twocolumn,amsmath,amssymb,groupaddress,longbibliography,eqsecnum]{revtex4-1}
\usepackage{graphicx,amsmath,relsize,epstopdf,upgreek,color,mathtools,bm,mathptmx}
\usepackage[hyphenbreaks]{breakurl}

\newcommand{\ket}[1]{\left|{#1}\right>}
\newcommand{\bra}[1]{\left<{#1}\right|}

\newcommand{\opinner}[3]{\left<{#1}\vphantom{#1}\vphantom{#3}\right|{#2}\left|{#3}\vphantom{#1}\vphantom{#3}\right>}
\newcommand{\tr}[1]{\mathrm{tr}\!\left\{#1\right\}}

\newcommand{\I}{\mathrm{i}}

\newcommand{\E}[1]{\mathrm{e}^{\mbox{\footnotesize$#1$}}}

\newcommand{\ML}{\widehat{\rho}_\textsc{ml}}

\catcode`,\active

\catcode`\,12

\begin{document}

\title{Adaptive compressive tomography with no \emph{a priori} information}

\author{D.~Ahn}
\affiliation{Department of Physics and Astronomy, Seoul National University, 08826 Seoul, South Korea}

\author{Y.~S.~Teo}
\email{ys\_teo@snu.ac.kr}
\affiliation{Department of Physics and Astronomy, Seoul National University, 08826 Seoul, South Korea}

\author{H.~Jeong}
\affiliation{Department of Physics and Astronomy, Seoul National University, 08826 Seoul, South Korea}

\author{F.~Bouchard}
\affiliation{Physics Department, University of Ottawa,
  25 Templeton Street, Ottawa, Ontario, K1N 6N5 Canada}

\author{F.~Hufnagel}
\affiliation{Physics Department, University of Ottawa,
  25 Templeton Street, Ottawa, Ontario, K1N 6N5 Canada}

\author{E.~Karimi}
\affiliation{Physics Department, University of Ottawa,
  25 Templeton Street, Ottawa, Ontario, K1N 6N5 Canada}

\author{D.~Koutn{\'y}}
\affiliation{Department of Optics, Palack\'{y}  University,
	17. listopadu 12, 77146 Olomouc, Czech Republic}

\author{J.~\v{R}eh\'{a}\v{c}ek}
\affiliation{Department of Optics, Palack\'{y}  University,
	17. listopadu 12, 77146 Olomouc, Czech Republic}

\author{Z.~Hradil}
\affiliation{Department of Optics, Palack\'{y}  University,
	17. listopadu 12, 77146 Olomouc, Czech Republic}

\author{G.~Leuchs}
\affiliation{Max-Planck-Institut f\"ur  die Physik des Lichts,
	Staudtstra\ss e 2, 91058 Erlangen, Germany}

\author{L.~L.~S\'{a}nchez-Soto}
\affiliation{Max-Planck-Institut f\"ur  die Physik des Lichts,
	Staudtstra\ss e 2, 91058 Erlangen, Germany}

      \begin{abstract}
	Quantum state tomography is both a
	crucial component in the field of quantum information and
	computation, and a formidable task that requires an
	incogitable number of measurement configurations as the
	system dimension grows. We propose and experimentally carry
	out an intuitive adaptive compressive tomography scheme,
	inspired by the traditional compressed-sensing protocol in
	signal recovery, that tremendously reduces the number of
	configurations needed to uniquely reconstruct any given
	quantum state without any additional \emph{a priori}
	assumption whatsoever (such as rank information, purity, etc)
	about the state, apart from its dimension.
\end{abstract}

\maketitle

\emph{Introduction.---}The characterization of an unknown (true) quantum state
$\rho_\text{t} \geq 0$ of Hilbert-space dimension $d$ is a subject of
immense study in quantum
information~\cite{Chuang:2000fk,lnp:2004uq,Teo:2015qs}. To fully
reconstruct an \emph{arbitrary} $\rho_\text{t}$, one may perform a set
of measurements that is enough to characterize all
$d^2-1$ independent parameters that define
$\rho_\text{t}$. Unfortunately, the number of such measurements
generally grows polynomially with $d$, or exponentially with the
number of subsystems that determine the quantum-source
complexity. This poses a technical limitation on how far conventional
quantum tomography can go in practical
experiments~\cite{Haffner:2005aa,Titchener:2018aa}.

If we know \emph{a priori} that $\mathrm{rank} \{\rho_\text{t}=\rho_{r}\}\leq r$ is extremely small, $r \ll d$, then the concept of
compressed sensing (CS), whose foundation was first mathematically
laid in the context of
imaging~\cite{Donoho:2006cs,Candes:2006cs,Candes:2009cs}, facilitates
the search for a unique estimator by measuring much fewer
configurations~\cite{Gross:2010cs,Kalev:2015aa,Steffens:2017cs,Riofrio:2017cs}. We
say that the corresponding data are \emph{informationally complete
	(IC)} for $\rho_{r}$. The state-of-the-art CS measurements to be
performed given such a prior information have been constructed
in~\cite{Baldwin:2016cs}.

The standard CS procedure, nevertheless, has two important issues that
need to be addressed. Firstly, an \emph{a priori} knowledge about $r$
is necessary to establish a preliminary order-of-magnitude estimate
for the number of configurations needed to fully characterize $\rho_r$
of rank no larger than $r$. Accuracy of the final estimator is hence
highly dependent on the validity of this \emph{a priori}
guess. Secondly, one has no means of verifying whether
the measurement data at hand are truly IC for $\rho_r$ in the standard
scheme. Typically, accuracy surveys with target states are
employed~\cite{Kalev:2015aa,Steffens:2017cs,Riofrio:2017cs} and the
value of such a survey relies on the precision of these target
states. Therefore, the decision of \emph{a priori} rank information
and presumed choices of target states are ultimately debatable in the
presence of experimental errors, rendering the reliability of any
related tomography scheme questionable.

In this Letter, we establish a new adaptive tomography paradigm
that completely removes the need for any sort of \emph{a priori}
information about $\rho_r$ (except for its dimension $d$). Our
proposed \emph{adaptive compressive tomography}~\mbox{(ACT)}
also includes an efficient recipe to determine informational
completeness of the collected data. No target states are ever required
to validate the resulting state estimator. The convex boundary
of the quantum state space and the positivity constraint plays the
principal role in checking whether the accumulated data are IC and
adaptively choosing measurements efficiently to uniquely reconstruct
$\rho_r$, the two of which completely define the purpose of ACT.

To demonstrate ACT, we perform an experiment with the
orbital angular momentum (OAM) of single photons and apply ACT to
states of various ranks engineered in these degrees of
freedom. Both experimental and simulated results show that ACT
requires a smaller number of measurements to
reconstruct rank-deficient quantum states as compared
to conventional CS tomography with known types of CS measurements.

%%%%%%%%%%%%%%%%%%%%%%%%%%
\begin{figure}[htp]
	\centering
	\includegraphics[width=.90\columnwidth]{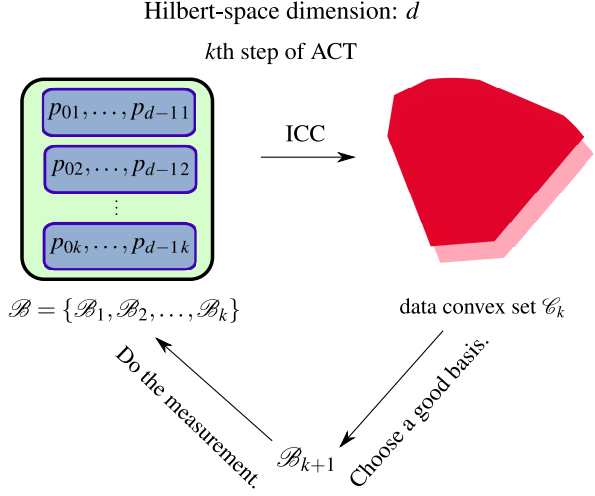}
	\caption{\label{fig:schema}Schematic diagram of a
		particular adaptive step in ACT tomography. In a clockwise flow,
		ACT first performs ICC to check whether data (blue) collected from
		measuring $\mathcal{B}$ are IC or not. If not, it proceeds to choose
		a good basis to measure in the next step.}
\end{figure}
%%%%%%%%%%%%%%%%%%%%%%%%%

\emph{The quantum state space and ACT.---}In the absence of statistical fluctuations, we measure a randomly chosen computational basis
$\{\ket{0},\ket{1},\ldots,\ket{d-1} \}$ on $\rho_r$ of
Hilbert-space dimension $d$. The corresponding Born
probabilities $p_j=\opinner{j}{\rho_r}{j}$ $(0\leq j\leq d-1)$
specify only the diagonal elements of $\rho_r$, and there is in
principle a \emph{data convex set}
$\mathcal{C}=\{\rho|\rho\leftrightarrow p_j\,\,\forall j\}$ comprising
infinitely many estimators $\widehat{\rho}$ that are consistent with
$p_j$. Evidently, $\rho_r\in\mathcal{C}$, and so the only fundamental
objective of ACT is to shrink $\mathcal{C}$ to a single point with
only $k_\textsc{ic} \ll d+1$ IC measurement bases for
$r\ll d$. For noiseless situations, this point must be
$\rho_r$.

To gain insights into how quantum positivity constraint plays a major
role in shrinking $\mathcal{C}$, we argue
in Appendix~\ref{sec:app1}
that if one methodically measures $k_0=\lceil(r^2-r)/(d-1)\rceil+1$
orthonormal bases, one of which being the eigenbasis
$\mathcal{B}_{\rho_r}$ of $\rho_r$, then $\widehat{\rho}=\rho_r$ is
the \emph{unique positive estimator} consistent with all measured
probabilities. For $r\ll d$, the regime of our interest, $k_0=2$. It
is however clear that $k_\textsc{ic}>2$ in real-world settings where
$\rho_r$ is completely unknown (apart from its dimension $d$), so the
famous no-go answer to Pauli's phase-retrieval
problem~\cite{Pauli:1933pr,Carmeli:2015cs} still stands. Regardless
the positivity constraint can still ensure an efficient compression of
the IC tomography procedure solely by data analysis.

The goal of ACT is to uniquely reconstruct any given unknown
$\rho_r$ through adaptively measuring one orthonormal basis at a time
according to collected data, as sketched in
Fig.~\ref{fig:schema}. In the $k$th step of the adaptive scheme, ACT
performs two main procedures. (I)~First, ACT checks whether the
probabilities $p_{j'k'}=\tr{\rho_r\Pi_{j'k'}}$ obtained from outcomes
$\Pi_{j'k'}>0$ $\left(\Pi_{j'k'}\Pi_{j''k'}=\delta_{j',j''}\right)$ of
all the measured orthonormal bases
$\mathcal{B}=\{\mathcal{B}_1,\mathcal{B}_2,\ldots,\mathcal{B}_k\}=\{{\Pi_{01},\ldots,\Pi_{d-1\,\,1}},\ldots,{\Pi_{0k},\ldots,\Pi_{d-1\,\,k}}\}$
so far are IC. Since the \emph{accumulated} data define a data convex
set $\mathcal{C}_k$ of size $s_k$ that contains all quantum states
$\rho$ consistent with $p_{j'k'}$, this procedure is tantamount to
finding out whether $s_k$ is zero or not. If $s_{k=k_\textsc{ic}}=0$,
then the estimator $\widehat{\rho}_{k=k_\textsc{ic}}\geq0$ consistent
with the IC data is unique by definition, and equal to $\rho_r$ when
statistical fluctuation is absent. (II)~If $s_k\neq0$, the accumulated
data collected are not IC and ACT shall choose the next basis by
analyzing $\mathcal{C}_k$. Beginning with $k=1$, a ``good'' adaptive
bases sequence should lead to a quick convergence of
$\mathcal{C}\rightarrow\rho_r$ as ACT progresses.

%%%%%%%%%%%%%%%%%%%%%%%%%%%
\begin{figure}[t]
	\centering
	\includegraphics[width=1\columnwidth]{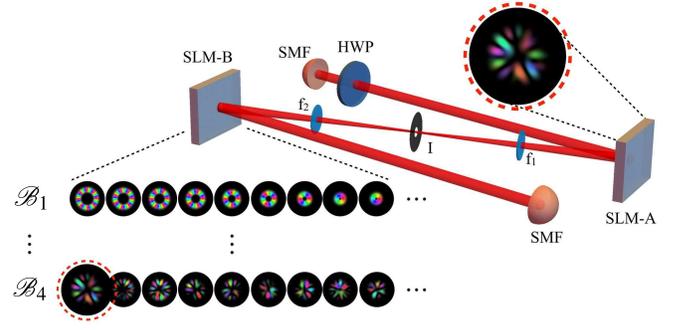}
	\caption{\label{fig:expt}
		Schematic of the OAM-based experimental setup.
		A 16-dimensional OAM state  is generated at SLM-A using a holographic
		technique that allows the tailoring of the intensity  and phase profile of the
		incoming beam. The modulated first-order of diffraction is filtered out using
		an iris (I) and a pair of lenses ($\text{f}_1$ and $\text{f}_2$).
		A similar holographic technique is used at the second SLM-B to measure
		the state in a given basis. The first measurement basis, $\mathcal{B}_1$,
		is given by the OAM computational basis. In the case of the rank 1 state shown
		on SLM-A, the corresponding eigenbasis is achieved after the fourth iteration.}
\end{figure}
%%%%%%%%%%%%%%%%%%%%%

\emph{(I)~Informational completeness certification (ICC).---}To certify whether all collected data are IC or not in the $k$th
adaptive step, it suffices to note that since $\mathcal{C}_k$ is
convex, maximizing and minimizing the linear function
$f_Z(\rho)=\tr{\rho Z}$ for some operator $Z$ over
$\rho\in\mathcal{C}_k$ respectively give unique solutions
$\rho_\text{max}$ and $\rho_\text{min}$ to the corresponding maximum
$f_{\text{max},k}$ and minimum $f_{\text{min},k}$. Without loss of
generality, $Z$ is taken to be a random full-rank state. We may define the
quantity
$s_{\textsc{cvx},k}=(f_{\text{max},k}-f_{\text{min},k})/(f_{\text{max},1}-f_{\text{min},1})$
that is a \emph{size monotone} (see Appendix~\ref{sec:app2}) for $\mathcal{C}_k$ in the sense that
$s_k<s_{k-1}$ if $s_{\textsc{cvx},k}<s_{\textsc{cvx},k-1}$---it is a
witness for the shrinkage of $\mathcal{C}_k$. As more
linearly independent bases are measured,
$s_{\textsc{cvx},k}\geq s_{\textsc{cvx},k+1}$, and
$s_{\textsc{cvx},k_\textsc{ic}}=0$ implies that $s_{k_\textsc{ic}}=0$
and that all data collected are IC for a unique reconstruction of
$\rho_r$. Therefore, at every adaptive step in ACT, we run:
%%%%%%%%%%%%%%%%%%%%%%%%%%%%%%%
\begin{center}
	\begin{minipage}[c][6cm][c]{0.9\columnwidth}
		\noindent
		\rule{\columnwidth}{1.5pt}\\
		\textbf{ICC in the $k$th step}
		\begin{enumerate}
			\item Maximize and minimize $f_Z(\rho)=\tr{\rho Z}$ for a fixed, randomly-chosen full-rank state $Z\neq1/d$ to obtain $f_{\text{max},k}$ and $f_{\text{min},k}$
			subject to
			\begin{itemize}
				\item $\rho\geq0$\,, $\tr{\rho}=1$\,,
				\item $\tr{\rho\Pi_{j'k'}}=p^{(\textsc{ml})}_{j'k'}$ for $0\leq j'\leq d-1$ and $1\leq k'\leq k$\,.
			\end{itemize}
			\item Compute $0\leq s_{\textsc{cvx},k}\leq1$ and check if it is smaller than some threshold $\epsilon$.
			\item If $s_{\textsc{cvx},k}<\epsilon$, terminate ACT. Continue otherwise.\\[-6ex]
		\end{enumerate}
		\rule{\columnwidth}{1.5pt}
	\end{minipage}
\end{center}
%%%%%%%%%%%%%%%%%%%%%%%%%%%%%%

The aforementioned strategy is, as a matter of fact, a semidefinite program (SDP)~\cite{Vandenberghe:1996sd} that can be efficiently solved by a variety of numerical methods. We should clarify here that while determining whether a set of measurement bases $\mathcal{B}$ possesses the conventional CS property for the entire class of rank-$r$ states is an NP-hard problem~\cite{Bandeira:2013aa}, ascertaining whether $\mathcal{B}$ gives a unique estimator for one unknown $\rho_r$ with the measurement data is, on the other hand, only as computationally difficult as carrying out the semidefinite program in ICC with a worst-case polynomial complexity~\cite{Vandenberghe:1996sd}.

For experimental data $\sum_{j'}\nu_{j'k'}=1$ ($1\leq k'\leq k$) with statistical noise, $\mathcal{C}_k$ is defined as the \emph{maximum-likelihood (ML) convex set} in which all $\rho\in\mathcal{C}_k$ satisfy the physical constraints $p^{(\textsc{ml})}_{j'k'}=\tr{\rho\Pi_{j'k'}}$ imposed by the ML principle for quantum states $\left[p^{(\textsc{ml})}_{j'k'}\rightarrow p_{j'k'}\,\text{ for }\,N\rightarrow\infty\right]$~\cite{lnp:2004uq,Teo:2015qs,Rehacek:2007ml,Shang:2017sf}. All arguments for noiseless data hold exactly for the ML probabilities, so that the working principle of ICC is \emph{perfectly robust against arbitrary noise} in the sense that $s_{\textsc{cvx},k_\textsc{ic}}=0$ still implies  $s_{k_\textsc{ic}}=0$ for noisy data owing to the preserved convexity of the newly defined $\mathcal{C}_k$. Noise only affects the reconstruction accuracy of the final unique estimator relative to $\rho_r$, which is a different subject matter for discussion.

\emph{(II)~Adaptive selection of measurement bases.---}The
optimal orthonormal basis to pick in the $k$th step and measure
in the $(k+1)$th step is the one that minimizes
$s_{\textsc{cvx},k+1}$. Since $\rho_r$ is unknown, we can treat some
\emph{a posteriori} estimator $\widehat{\rho}_k$ from $\mathcal{C}_k$
as a guess for $\rho_r$ to generate simulated data during the
minimization of $s_{\textsc{cvx},k+1}$ over all future basis
choices. The complicated dependence of $s_{\textsc{cvx},k+1}$ on the
future basis however makes its brute-force optimization
computationally exhaustive for large $d$.

For a more tractable approach to adaptively measure good bases,
we first note that $\mathcal{C}_{k<k_\textsc{ic}}$ \emph{essentially} contains states with eigenbases that are distinct from $\{\mathcal{B}_1\ldots\mathcal{B}_k\}$ (see Appendix~\ref{sec:app3}). So even if we know nothing
about $\rho_r$, if it is rank-deficient, then taking
$\mathcal{B}_{k+1}$ to be the diagonal basis of a rank-deficient
$\widehat{\rho}_k\in\mathcal{C}_k$ ensures a distinct measurement basis
in each step that generates a reasonably fast converging sequence $\mathcal{B}_{k}\rightarrow\mathcal{B}_{\rho_r}$ as $k$ increases since $\mathcal{C}_k\rightarrow\rho_r$ at the same time. There is more than one approach to pick eigenbases of rank-deficient states from $\mathcal{C}_k$, and as an example we shall consider the minimization of von Neumann entropy function $S(\rho)=-\tr{\rho\log\rho}$. A superfast algorithm suitable for minimizing $S$ over $\mathcal{C}_k$ exists~\cite{Shang:2017sf,Teo:2011me}. Incidentally, it was reported in \cite{Huang:2016aa,Tran:2016aa} that entropy minimization also offers high compressive efficiencies in both sparse-signal and low-rank matrix recovery.

\emph{Complete ACT protocol.---}All aforementioned arguments can
accommodate real experimental situations, where the relative frequency
data do not typically correspond to physical quantum states for
$k>1$. The data convex sets contain states that are now consistent
with the corresponding physical ML probabilities derived from data,
which are statistically consistent with the true probabilities. The
final unique estimator $\widehat{\rho}_{k_\textsc{ic}}$ would then
incur a statistical bias from $\rho_r$ that drops as $N$
increases. For many-body quantum sources, the bases generated by ACT are entangled. In practice, product bases are typically much more practical to implement for such sources. While verifying if a rank-deficient $\widehat{\rho}_k\in\mathcal{C}_k$ can possess a product eigenbasis is computationally difficult, ACT can still be adjusted to feasibly generate near-optimal product bases~(pACT) by defining $\mathcal{B}_{k+1}$ to be the product basis that is nearest to the eigenbasis of $\widehat{\rho}_k$ with respect to some given norm using a nonlinear optimization routine. Both ACT and pACT for any experimental setting are summarized:
\begin{center}
	\begin{minipage}[c][11cm][c]{0.9\columnwidth}
		\noindent
		\rule{\columnwidth}{1.5pt}\\
		\textbf{ACT/pACT}\\
		\noindent
		Beginning with $k=1$ and a random computational basis $\mathcal{B}_1$:
		\begin{enumerate}
			\item Measure $\mathcal{B}_k$ and collect the relative frequency data $\sum^{d-1}_{j'=0}\nu_{j'k}=1$.
			\item From $\left\{\nu_{0k'},\ldots,\nu_{d-1\,\,k'}\right\}^k_{k'=1}$, obtain $kd$ physical ML probabilities.
			\item Perform ICC with the ML probabilities and compute $s_{\textsc{cvx},k}$:
			\begin{itemize}
				\item \textbf{If}~$s_{\textsc{cvx},k}<\epsilon$, terminate ACT and take $\rho_\text{max}\approx\rho_\text{min}$ as the estimator and report $s_{\textsc{cvx},k}$.
				\item \textbf{Else}~Proceed.
			\end{itemize}
			\item Choose a rank deficient $\widehat{\rho}_k\in\mathcal{C}_k$ [for instance by minimizing the von Neumann entropy $S(\rho)$ in $\mathcal{C}_k$].
			\item Define $\mathcal{B}_{k+1}$ to be the eigenbasis of $\widehat{\rho}_k$ for ACT, or a basis close to it for pACT \emph{via} some prechosen distance minimization technique.
			\item Set $k=k+1$ and repeat.\\[-6ex]
		\end{enumerate}
		\rule{\columnwidth}{1.5pt}
	\end{minipage}
\end{center}

\emph{Analysis and experiments.---}We put both ACT and pACT tomography schemes to the experimental test by comparing their results with those from measuring random Pauli (RP) bases considered in \cite{Kalev:2015aa,Steffens:2017cs,Riofrio:2017cs}, the Baldwin--Goyeneche~(BG) bases in \cite{Baldwin:2016cs} that generalizes a known five-bases construction for $r=1$ to an IC set of $k_\textsc{ic}=4r+1$ bases for rank-$r$ quantum states, and the set of random orthonormal bases of $k_\textsc{ic}\approx\lceil4r(d-r)/(d-1)\rceil$ studied in \cite{Kech:2016aa}. This exact scaling shall be used to benchmark the experimental $k_\textsc{ic}$s.

To demonstrate all three schemes (see
Fig.~\ref{fig:expt}), we experimentally emulate a 4-qubit ($d=16$) quantum system and both entangled and product measurement bases using an OAM-based setup. In particular, we consider the
Laguerre-Gauss (LG) modes with azimuthal and radial mode indices
$\ell$ and $p=0$, respectively. Hence, OAM states correspond to a
sub-space of the LG modes and are characterized by a helical
wavefront given by $\E{\I\ell\phi}$, where $\ell$ is the
azimuthal index that corresponds to the OAM value, and $\phi$ is
the azimuthal coordinate. The appropriate phase and intensity
patterns are realized using a holographic technique called
\textit{intensity masking}, which is readily achieved by a
programmable spatial light modulator (SLM)~\cite{Bolduc:2013aa}. By doing so, we can prepare \emph{any} many-body state and measurement basis. The
generated photons are detected using the projective technique of \textit{intensity-flattening}~\cite{Bouchard:2018aa}, where any
arbitrary spatial mode can be measured using an SLM followed by a
single mode fiber (SMF).

\begin{figure}[t]
	\centering
	\includegraphics[width=1\columnwidth]{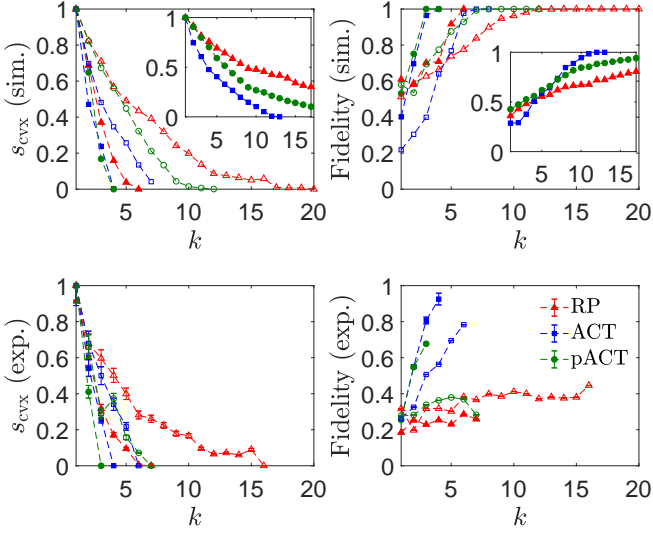}
	\caption{\label{fig:size_dist}Plots of simulation (noiseless) and experimental values of $s_\textsc{cvx}$ and $\widehat{\rho}_k$ fidelity against the measured basis number $k$ for $d=16$, where $\widehat{\rho}_k:=\rho_\text{min}$ for the RP scheme. Estimated experimental error bars reflect the propagated Poissonian-source standard deviations. All plot markers are averaged over five $\rho_r$s. The filled markers represent results for rank-1 $\rho_r$s whereas the unfilled ones represent those for rank-3 $\rho_r$s. The insets showcase simulation performances of rank-6 states as a demonstration of high-rank $(r\approx D/2)$ compressive tomography, with $1\leq k\leq(D+1)=17$ restricted to the minimal bases number for arbitrary-state tomography. The lower experimental fidelities for RP and pACT are due to a technical bias of the OAM setup for finite $N$, where bases close to the eigenbasis of $\rho_r$ tend to give estimated Born probabilities that are relatively more accurate than those that are not. So for OAM sources, ACT is the most favorable option, as both pACT and RP correspond to measurement bases that are never close to the eigenbasis of $\rho_r$. Even with noisy data, ICC can still validate whether the resulting ML probabilities obtained from data are IC (left panels), which is the point of ACT.}
\end{figure}

A heralded single photon source is achieved by
pumping a 3~mm $\beta$-barium borate type I nonlinear crystal with a
quasi-continuous wave laser at a wavelength of 355~nm, producing
photon pairs at 710~nm via spontaneous parametric down-conversion. A
coincidence rate of 40~kHz, within a coincidence time window of
5~ns, is measured after filtering the photons to the fundamental
Gaussian modes using SMF. Subsequent to the generation and detection
of the photonic states, explained above, coincidence measurements
are recorded using single photon detectors and a coincidence logic.

\begin{figure}[t]
	\centering
	\includegraphics[width=1\columnwidth]{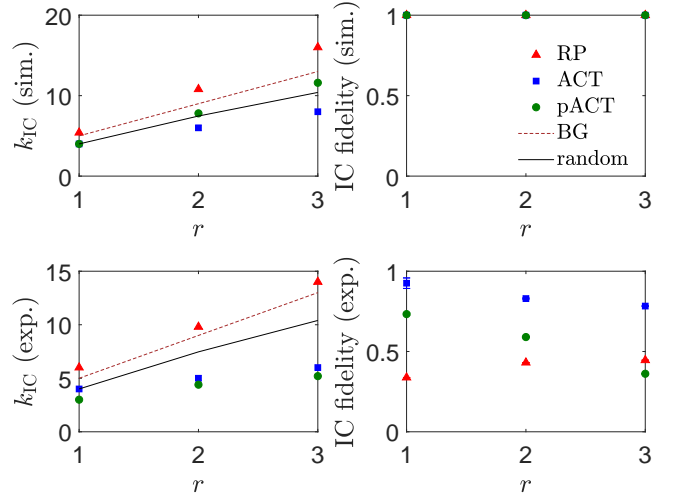}
	\caption{\label{fig:kic}Plots of simulation (noiseless) and experimental values of $k_\textsc{ic}$ and the $\widehat{\rho}_{k=k_\textsc{ic}}$ fidelity against the rank $1\leq r\leq3$ of $\rho_r$. All $k_\textsc{ic}$ values ascertained using ICC are averaged over five $\rho_t$s per rank. Otherwise all specifications are the same as Fig.~\ref{fig:size_dist}. Although for real data, positivity modifies the $k_\textsc{ic}$ performances with ML, pACT achieves informational completeness much quicker than RP as far as local bases are concerned. A comparison with random IC orthonormal bases shows that ACT gives a much lower value owing to the additional assessment of and optimization over $\mathcal{C}$.}
\end{figure}

All results are summarized in Figs.~\ref{fig:size_dist} and
\ref{fig:kic}, and the messages conveyed are succinctly stated here: For noiseless simulated data, in terms of \emph{average} $k_\textsc{ic}$ over
uniformly (Hilbert-Schmidt) distributed rank-$r$ true states, ACT is
the most efficient, since it guides the measurement basis to the
eigenbasis of $\rho_r$. The more many-body-suited pACT that
adaptively generates product bases requires a larger $k_\textsc{ic}$
to yield IC data, but the average performance margin with ACT is
narrow for low-$r$ states and is on par with the scaling of entangled
Goyeneche-type bases ($k_\textsc{ic}=4r+1$) for larger $r$. RP turns
out to be least efficient amongst all tested schemes. Even in the
presence of real data noise, both ACT and pACT remain the more
favorable candidates for tomography on general complex systems.

\emph{Concluding remarks.---}The feasible concept of adaptive compressive tomography developed here
provides a powerful method to reconstruct any unknown rank-deficient
quantum state with optimally chosen entangled or product orthonormal
measurement bases, especially for quantum sources of complex degrees
of freedom, which includes many-body systems. More importantly, the
adaptive scheme requires no \emph{a priori} knowledge or assumptions
about the state or near-proximity target states because it can
self-sufficiently validate whether the measured data are
informationally complete or not using semidefinite programming, so
that reliable compressive tomography can now be carried out in real
experimental situations with noisy data. The superior compressive
efficiencies of both entangled and product versions of our adaptive
schemes are confirmed experimentally and demonstrated with respect to
other established protocols.

We acknowledge financial support from the BK21 Plus Program
(21A20131111123) funded by the Ministry of Education (MOE, Korea) and
National Research Foundation of Korea (NRF), the NRF grant funded by
the Korea government (MSIP) (Grant No. 2010-0018295), the Korea
Institute of Science and Technology Institutional Program (Project
No. 2E27800-18-P043), the European Union's Horizon 2020 Research
and Innovation Programme under Grant Agreement No. 766970, Canada
Research Chairs (CRC), the Spanish MINECO (Grant
No. FIS2015-67963-P), the Grant Agency of the Czech Republic (Grant
No. Grant No. 18-04291S), and the IGA Project of the
Palack{\'y} University (Grant No. IGA PrF 2018-003).

\appendix
\section{Uniqueness property induced by the state-space boundary}
\label{sec:app1}

Suppose that the eigenbasis $\{\ket{\lambda_j}\}$ of a rank-$r$ 
\begin{equation}
\rho=\rho_r=\sum^{r-1}_{j=0}\ket{\lambda_j}p_j\bra{\lambda_j}
\end{equation}
is measured and the probabilities $\sum^{r-1}_{j=0}p_j=1$ are collected, the estimator $\widehat{\rho}=\rho_r$ is one solution that is consistent with the Born probabilities $\opinner{\lambda_j}{\widehat{\rho}}{\lambda_j}=p_j$. If $\widehat{\rho}$ is only required to be Hermitian, then the general solution is in fact given by 
\begin{equation}
\widehat{\rho}=\rho_r+\sum^{r-1}_{j\neq k=0}\ket{\lambda_j}c_{jk}\bra{\lambda_k}+W\,,
\label{eq:rhoest}
\end{equation}
where $W$ is a traceless Hermitian operator outside the support of $\rho$ ($W\rho=0=\rho W$). Moreover, $p_{j\geq r}=0=\opinner{\lambda_{j\geq r}}{\widehat{\rho}}{\lambda_{j\geq r}}$ implies that $W$ is represented by a hollow (all diagonal entries equal to zero) Hermitian matrix in the basis $\{\ket{\lambda_{j\geq r}}\}$ with arbitrary off-diagonal entries. The Hermitian solution subspace for $\widehat{\rho}$ has thus a nonzero volume under some metric. 

It is now obvious that if one chooses $\ket{\phi}$ to be the eigenket of $W$ that gives a negative eigenvalue, one expects $\opinner{\phi}{\widehat{\rho}}{\phi}=\opinner{\phi}{W}{\phi}<0$. This implies that any such nonzero traceless $W$ always results in a nonpositive $\widehat{\rho}$. If the quantum positivity constraint is imposed on the solution for the Born probabilities, we must necessarily have $W=0$. 

This leaves the operator $A=\sum^{r-1}_{j\neq k=0}\ket{\lambda_j}c_{jk}\bra{\lambda_k}$ in the right-hand side of Eq.~\eqref{eq:rhoest}. For pure states $(r=1)$, $A$ is clearly zero, so that $\widehat{\rho}=\rho_1=\ket{\lambda_1}\bra{\lambda_1}$ is the unique positive estimator after measuring $\rho_1$ as we expect. For $1<r\leq d$, we note that if $\widehat{\rho}$ is to be consistent with probabilities $\opinner{w_j}{\widehat{\rho}}{w_j}=\opinner{w_j}{\rho_r}{w_j}$ obtained from any other orthonormal measurement basis $\{\ket{w_j}\}$, then $\opinner{w_j}{A}{w_j}=0$ for all $1\leq j\leq d-1$. As $A$ has exactly $r^2-r$ free parameters, it follows that measuring $k_0=\lceil(r^2-r)/(d-1)\rceil+1$ linearly independent bases results in the unique solution $A=0$ to $k_0(d-1)$ \mbox{(pseudo-)}invertible linear equations. When $r^2-r\leq d-1$, measuring just one basis other than the eigenbasis will result in a unique $\widehat{\rho}$. If $r=d$, we then obtain the familiar minimal bases number $k_0=d+1$. 

To summarize, the quantum positivity constraint restricts the possible solution set consistent with probabilities derived from any given rank-$r$ state $\rho_r$ to a unique state estimator if the eigenbasis of $\rho_r$ and $\lceil(r^2-r)/(d-1)\rceil$ other orthonormal bases are measured. The method of ACT strives to achieve a small number $k_\textsc{ic}$ of optimal IC bases that is bounded from below by $k_0$. Looking at two extremal cases, measuring $\mathcal{B}_{\rho_r}$ immediately gives us the unique estimator for any pure state, whereas the minimal number of bases needed to characterize a full-rank state is certainly $k_0=d+1$.

\section{The size monotone}
\label{sec:app2}

The \emph{size monotone} $0\leq s_{\textsc{cvx},k}\leq1$ for the data convex set $\mathcal{C}_k$ is a (non-strict) monotonically increasing function with its size $s_k$ ($s_{\textsc{cvx},k}>s_{\textsc{cvx},k+1}\implies s_k>s_{k+1}$).

To define a size monotone for any data convex set $\mathcal{C}_k$, we first pre-choose a concave function $f(\rho)$ of a unique maximum to characterize $\mathcal{C}_k$. In the absence of statistical fluctuation, we have the set inequality chain $\mathcal{C}_1\supseteq\mathcal{C}_2\supseteq\ldots\supseteq\mathcal{C}_{d+1}$ as linear independent bases are sequentially measured: states in the data convex set are ruled out as more information is gained through measurements. By the concavity of $f(\rho)$ we have $f_{\text{max},1}\geq f_{\text{max},2}\geq\ldots\geq f_{\text{max},d+1}$ and $f_{\text{min},1}\leq f_{\text{min},2}\leq\ldots\leq f_{\text{min},d+1}$. It is clear that if $f_{\text{max},k+1}-f_{\text{min},k+1}<f_{\text{max},k}-f_{\text{min},k}$, then $\mathcal{C}_{k+1}\subset\mathcal{C}_{k}$ or $s_{k+1}<s_k$.

It follows immediately that if $s_{\textsc{cvx},k}\equiv(f_{\text{max},k}-f_{\text{min},k})/(f_{\text{max},1}-f_{\text{min},1})$, then $s_{\textsc{cvx},k}$ is a size monotone that decreases with increasing $k$. When $s_{\textsc{cvx},k=k_\textsc{ic}}=0$, the convexity of $\mathcal{C}_{k_\textsc{ic}}$ implies that $s_{k_\textsc{ic}}=0$ as $\mathcal{C}_{k_\textsc{ic}}$ must contain only $\rho_r$ due to the unique maximum possessed by $f$. Similar arguments hold for a convex $f$. Since $f_Z(\rho)$ is a positive linear function it can also be used to formulate the size monotone as it facilitates the class of semidefinite programs known to give a unique maximum as well as minimum in the quantum state space.

It is easy to see that $s_{\textsc{cvx},k_\textsc{ic}}=0$ still implies that $s_{k_\textsc{ic}}=0$ with noisy data that may come from statistical fluctuation or other systematic errors. For this, we can define $\mathcal{C}_k$ to be the set of states that maximizes the likelihood function, of the exemplifying multinomial form $L(n_{j'k'}|\rho')=\prod^{k}_{k'=1}\prod^{d-1}_{j'=0}{p'_{j'k'}}^{n_{j'k'}}$ for typical physical problems involving independent single-copy sampling up to a fixed total sample size $\sum^{k}_{k'=1}\sum^{d-1}_{j'=0}n_{j',k'}=N$ derived from the observed frequencies $n_{j'k'}$ labeled by the outcome $j'$ and basis $k'$ numbers. It is clear that if the data are noiseless, the maximum of $L$ gives precisely $\ML=\rho_r$, so such a definition is a valid generalization to real experimental scenarios. It is important to note that while the set inequality chain ``$\mathcal{C}_{k+1}\subset\mathcal{C}_{k}$'' is in general broken, that is $s_{\textsc{cvx},k}$ no longer behaves as a size monotone in $k$, the newly defined $\mathcal{C}_k$s are still convex sets because $\log L(n_{j'k'}|\rho')$ is a concave function and hence possesses a convex plateau structure for non-IC data. The convexity of $\mathcal{C}_k$ arises more generally from any kind of concave $\log L(n_{j'k'}|\rho')$, which behavior is common in experiments. This is the only crucial property to again conclude that $s_{\textsc{cvx},k_\textsc{ic}}=0\implies s_{k_\textsc{ic}}=0$. It therefore follows that the ICC protocol introduced in the main article for ACT is perfectly robust to noisy data, in the sense that the set of measurement bases $\mathcal{B}=\{\mathcal{B}_1,\mathcal{B}_2,\ldots,\mathcal{B}_{k_\textsc{ic}}\}$, along with their collected data, always give a unique state reconstruction for the value of $k_\textsc{ic}$ decided by ICC even with noisy data. No premature termination of ACT will occur. On the other hand, the final unique state estimator, of course, will have a lower statistical accuracy because of data noise.

\section{The spectral decomposition of states in $\mathcal{C}$}
\label{sec:app3}

For an integer $k$ and a noiseless scenario, the data convex set $\mathcal{C}_k$ contains all quantum states that are consistent with the bases measurements $\mathcal{B}=\{\mathcal{B}_1\,\ldots,\mathcal{B}_k\}$. Suppose that $\mathcal{C}_k\neq\{\ML\}$ We would like to check how frequently a randomly chosen state $\rho'\in\mathcal{C}_k$ possesses an eigenbasis equal to $\mathcal{B}_j$ for some $1\leq j\leq k$. 

It is easy to see that if $\rho'=\sum_l\ket{\lambda_{lj}}\lambda_{lj}\bra{\lambda_{lj}}\in\mathcal{C}_k$ where $\mathcal{B}_j=\{\ket{\lambda_{lj}}\bra{\lambda_{lj}}\}$, then we must have the eigenvalues $\lambda_{lj}=p'_{lj}=\opinner{\lambda_{lj}}{\rho'}{\lambda_{lj}}$ according to the definition of $\mathcal{C}_k$. A trivial example occurs when $k=1$, where $\mathcal{C}_1$ contains exactly one diagonal state in the measurement basis. It follows immediately that there can exist at most $k$ states in $\mathcal{C}_k$ that possesses eigenbases overlapping with $\mathcal{B}$, which are clearly measure zero compared to the infinitely many states in $\mathcal{C}_k$. This implies that $\mathcal{C}_k$ for $k<k_\textsc{ic}$ contains states with eigenbases that are distinct from $\mathcal{B}$ as the only measurable states. We add that the actual number of states with such eigenbases is generally much lower than $k$ since for $k>1$, every state in $\mathcal{C}_k$ must satisfy \emph{all} probability constraints.

%\bibliography{Biblio}


%% just before \end{document}
%\cleardoublepage\newcommand{\twocolumn}{\relax}\newcommand{\onecolumn}{\relax}
%\input{LogLabs}\input{BibCheck}

\end{document}